\documentclass[a4paper,11pt]{article}
\pdfoutput=1 % if your are submitting a pdflatex (i.e. if you have
\usepackage{jheppub} % for details on the use of the package, please
\usepackage[T1]{fontenc} % if needed

\usepackage{slashbox}

\subheader{UAB-FT-755}

\title{\boldmath Combined analysis of the decays $\tau^-\to K_S\pi^-\nu_\tau$
and $\tau^-\to K^-\eta\nu_\tau$}

\author[a]{R.~Escribano,}
\author[a]{S.~Gonz\'alez-Sol\'is,}
\author[b]{M.~Jamin}
\author[c]{and P.~Roig}

\affiliation[a]{Grup de F\'{\i}sica Te\`orica (Departament de F\'{\i}sica) and
                Institut de F\'{\i}sica d'Altes Energies (IFAE), Universitat
                Aut\`onoma de Barcelona, E-08193 Bellaterra (Barcelona), Spain.}
\affiliation[b]{Instituci\'o Catalana de Recerca i Estudis Avan\c cats (ICREA),
                IFAE,\\ Universitat Aut\`onoma de Barcelona, E-08193 Bellaterra
                (Barcelona), Spain.}
\affiliation[c]{Instituto de F\'{\i}sica, Universidad Aut\'onoma de M\'exico,
                AP 20-364, M\'exico D.F. 01000, M\'exico.}

\emailAdd{rescriba@ifae.es}
\emailAdd{sgonzalez@ifae.es}
\emailAdd{jamin@ifae.es}
\emailAdd{pabloroig@fisica.unam.mx}

\abstract{In a combined study of the decay spectra of
$\tau^-\to K_S\pi^-\nu_\tau$ and $\tau^-\to K^-\eta\nu_\tau$ decays within a
dispersive representation of the required form factors, we illustrate how the
$K^*(1410)$ resonance parameters, defined through the pole position in the
complex plane, can be extracted with improved precision as compared to previous
studies. While we obtain a substantial improvement in the mass, the uncertainty
in the width is only slightly reduced, with the findings
$M_{K^{*\prime}}=1304 \pm 17\,$MeV and $\Gamma_{K^{*\prime}} = 171 \pm 62\,$MeV.
Further constraints on the width could result from updated analyses of the
$K\pi$ and/or $K\eta$ spectra using the full Belle-I data sample. Prospects
for Belle-II are also discussed. As the $K^-\pi^0$ vector form factor enters
the description of the decay $\tau^-\to K^-\eta\nu_\tau$, we are in a position
to investigate isospin violations in its parameters like the form factor
slopes. In this respect also making available the spectrum of the transition
$\tau^-\to K^-\pi^0\nu_\tau$ would be extremely useful, as it would allow to
study those isospin violations with much higher precision.}


\begin{document} 
\maketitle
\flushbottom

\section{Introduction}\label{Intro}

Hadronic decays of the $\tau$ lepton constitute a distinguished set of
processes to study the strong interactions in its non-perturbative regime
under rather clean conditions \cite{Braaten:1991qm,Braaten:1988hc,Narison:1988ni,Braaten:1988ea}.
This happens because the corresponding amplitudes can be factorised into a
purely electroweak part corresponding to the decay of the $\tau$ lepton into a
quark-antiquark pair and the associated $\tau$ neutrino, times the hadronization
of the left-handed quark bilinear current under the action of QCD. The
uncertainties of the first part are completely negligible with respect to those
of the second one, which allows a direct access to the hadronic currents that
has been exploited successfully for decades \cite{Pich:2013lsa}.

The dominant strangeness-changing $\tau$ decays are into $K\pi$ meson systems
and the corresponding observables have been measured with increasing precision
at LEP~\cite{aleph99,opal04}, BaBar~\cite{Aubert:2007jh} and
Belle~\cite{Epifanov:2007rf}. We would like to note that the BaBar collaboration
published their analysis for the $K^-\pi^0$ mode \cite{Aubert:2007jh}, while
Belle studied the $K_S\pi^-$ decay channel \cite{Epifanov:2007rf}. Belle's
spectrum became publicly available but the published BaBar analysis only
concerned the branching fraction while the corresponding spectrum has not
been released yet.\footnote{BaBar reported preliminary results for the
$\bar{K}^0\pi^-$ mode at the TAU'08 Conference \cite{Aubert:2008an}, whereas
Belle also plans to study the $K^-\pi^0$ mode and has just published updated
values of the branching fractions of decay modes including $K_S$ mesons
analysing a larger data sample \cite{Ryu:2014vpc}. We thank Swagato Banerjee,
Simon Eidelman, Denis Epifanov and Ian Nugent for conversations on this point.}
As a result, all dedicated studies of the $\tau^-\to (K\pi)^-\nu_\tau$ decays
focused on the $K_S\pi^-$ system \cite{Jamin:2006tk, Moussallam:2007qc, Jamin:2008qg, Boito:2008fq, Boito:2010me, Bernard:2013jxa}.
Consequently, even using data from semileptonic Kaon decays ($K \to \pi\ell\nu$,
so-called $K_{\ell 3}$ decays) \cite{Boito:2010me,Bernard:2013jxa}, important
information on isospin breaking effects in the low-energy expansion of the
hadronic form factors could not be extracted. The quoted references succeeded
in improving the determination of the $K^*(892)$ and $K^*(1410)$ resonance
properties: their pole positions and relative weight, although the errors on
the radial excitation were noticeably larger than in the $K^*(892)$
case.\footnote{Obviously, all these $\tau$-based analyses determined the
properties of the charged vector resonances. Those of the corresponding neutral
counterparts can only be accessed in meson-nucleon scattering or heavy flavour
decays, not in $e^+e^-$ experiments (where they are suppressed loop-mediated
effects). Since the theory input to analyse these is necessarily quite
different to that of hadronic $\tau$ decays, it is not easy to single out
isospin violations comparing the pole positions of both members of the
corresponding iso-doublets.}

The threshold for the decay $\tau^- \to K^-\eta\nu_\tau$ is above the region
of $K^*(892)$-dominance which enhances its sensitivity to the properties of
the heavier copy $K^*(1410)$. This observation was one of the motivations for
the analysis of ref.~\cite{Escribano:2013bca}, where it was first shown that
the considered decays were competitive to the $\tau^-\to (K\pi)^-\nu_\tau$
decays for the extraction of the $K^*(1410)$ meson parameters. This was made
possible thanks to BaBar \cite{delAmoSanchez:2010pc} and
Belle~\cite{Inami:2008ar} data of the $K^-\eta$ spectrum which improved
drastically the pioneering CLEO \cite{Bartelt:1996iv} and
ALEPH~\cite{Buskulic:1996qs} measurements.

The main purpose of this work is to illustrate the potential of a combined
analysis of the decays $\tau^- \to (K\pi)^-\nu_\tau$ and
$\tau^- \to K^-\eta\nu_\tau$ in the determination of the $K^*(1410)$ resonance
properties. This study is presently limited by three facts: unfolding of
detector effects has not been performed for the latter data, the associated
errors of these are still relatively large and no measurement of the $K^-\pi^0$
spectrum has been published by the B-factories. We intend to demonstrate that
an updated analysis of the $K_S\pi^-$ and/or $K^-\eta$ Belle spectrum including
the whole Belle-I data sample could improve notably the knowledge of the
$K^*(1410)$ pole position. Therefore, we hope that our paper strengths the
case for a (re)analysis of the $(K\pi)^-$ and $K^-\eta$ spectra at the first
generation B-factories including a larger data sample and also for devoted
analyses in the forthcoming Belle-II experiment. Turning to the low-energy
parameters, we emphasise the importance of (independent) measurements of
the two $\tau^-\to (K\pi)^-\nu_\tau$ charge channels with the target of
disentangling isospin violations in forthcoming studies.

While the $K\pi$-hadronization in the $\tau^-\to (K\pi)^-\nu_\tau$ decays is
quite well understood, earlier analyses of $\tau^-\to K^-\eta\nu_\tau$ decays
\cite{Pich:1987qq,Braaten:1989zn,Li:1996md} were at odds with Belle data (also
\cite{Kimura:2012} showed discrepancies) which motivated the claim in Belle's
paper \cite{Inami:2008ar} that `further detailed studies of the physical
dynamics in $\tau$ decays with $\eta$ mesons are required', as also observed
in ref.~\cite{Actis:2010gg} in a more general context. In
ref.~\cite{Escribano:2013bca}, we showed that a simple Breit-Wigner
parametrisation of the dominating vector form factor lead to a rather poor
description of the data, while more elaborated approaches based on Chiral
Perturbation Theory ($\chi PT$) \cite{Weinberg:1978kz,Gasser:1983yg,Gasser:1984gg}
including resonances as dynamical fields \cite{Ecker:1988te,Ecker:1989yg} and
resumming final-state interactions (FSI) encoded in the chiral loop functions
provided very good agreement with data. Since the $K^-\eta$ currents are
presently modelled in TAUOLA \cite{Jadach:1990mz, Jadach:1993hs} (the standard
Monte Carlo generator for $\tau$ lepton decays) relying on phase space, our
form factors will enrich the Resonance Chiral Lagrangian-based currents
\cite{Shekhovtsova:2012ra, Nugent:2013hxa} in the library (along these lines,
the inclusion of the dispersive treatment for the $K\pi$ system is also in
progress).

Our paper is organised as follows: in section \ref{Theory}, the differential
decay width of the $\tau^-\to K_S\pi^-/K^-\eta\,\nu_\tau$ processes is written
as a function of the contributing $K\pi$ vector and scalar form factors. The
vector form factors will be described according to a dispersive representation
along the lines of refs.~\cite{Boito:2008fq, Boito:2010me}, while the scalar
form factors are taken from refs.~\cite{JOP00,JOP02}, thereby resumming FSI
which is crucial to describe the considered decay spectra. Our previous
analysis of the $\tau^-\to K^-\eta\nu_\tau$ decays \cite{Escribano:2013bca}
disfavoured strongly the use of Breit-Wigner functions, both from the
theoretical and phenomenological perspective. In section \ref{Fits}, we
describe our fits in detail and present the corresponding results for all
parameters.  It will be seen that we are able to improve the determination of
the $K^*(1410)$ pole position. Furthermore, we discuss isospin violations on
the slope parameters of the vector form factors and the prospects for improving
them by analysing the full Belle-I data set or future measurements at Belle-II.
Finally, we summarise our conclusions in section \ref{Concl}. A brief
discussion of another so-called ``exponential'' parametrisation of the
$K\pi$ vector form factor which was put forward in
refs.~\cite{Jamin:2006tk,Jamin:2008qg} is relegated to Appendix~\ref{AppA}.

\section{Form factor representations}\label{Theory}

The differential decay width of the transition $\tau^-\to K_S\pi^-\nu_\tau$ as
a function of the invariant mass of the two-meson system can be written as
\begin{eqnarray}
\label{spectral function}
\frac{{\rm d}\Gamma(\tau^-\to K_S\pi^-\nu_\tau)}{{\rm d}\sqrt{s}} &\,=\,&
\frac{G_F^2 M_\tau^3}{96\pi^3s} S_{EW}\Big|V_{us}f_+^{K_S\pi^-}(0)\Big|^2
\biggl(1-\frac{s}{M_\tau^2}\biggr)^2q_{K_S\pi^-}(s) \\
\vbox{\vskip 8mm}
&\times& \biggl\{\left(1+\frac{2s}{M_\tau^2}\right)q_{K_S\pi^-}^2(s)\Big|\widetilde{f}_+^{K_S\pi^-}(s)\Big|^2+\frac{3\Delta_{K_S\pi^-}^2}{4s}
\Big|\widetilde{f}_0^{K_S\pi^-}(s)\Big|^2\biggr\} \,, \nonumber
\end{eqnarray}
where
\begin{equation}
\label{definitions}
q_{PQ}(s) \,=\, \frac{\sqrt{s^2-2s\Sigma_{PQ}+\Delta_{PQ}^2}}{2\sqrt{s}}\,,\quad
\Sigma_{PQ} \,=\, m_P^2+m_Q^2 \,, \quad
\Delta_{PQ} \,=\, m_P^2-m_Q^2 \,,
\end{equation}
and
\begin{equation}
\widetilde{f}_{+,0}^{PQ}(s) \,\equiv\, \frac{f_{+,0}^{PQ}(s)}{f_{+,0}^{PQ}(0)}
\end{equation}
are form factors normalised to unity at the origin. In this way, besides the
global normalisation, all remaining uncertainties on the hadronization of the
considered currents are encoded in the reduced form factors
$\widetilde{f}_{+,0}^{PQ}(s)$. $S_{EW} = 1.0201$ \cite{Erler:2002mv} resums
the short-distance electroweak corrections.\footnote{We have not included
additional non-factorisable electromagnetic corrections. They have been
estimated in ref.~\cite{Antonelli:2013usa} where it was found that at the
current level of precision they can be safely neglected.}
Eq.~(\ref{spectral function}) corresponds to the definitions of the vector,
$f_+^{PQ}(s)$, and scalar, $f_0^{PQ}(s)$, form factors that separate the P- and
S-wave contributions according to the conventions of ref.~\cite{Gasser:1984ux}.
The corresponding formula for the $\tau^-\to K^-\eta\nu_\tau$ decays can be
obtained by multiplying eq.~(\ref{spectral function}) with the ratio between
the corresponding SU(3) Clebsch-Gordan coefficients (three in this case) and
replacing the $K_S$ and $\pi^-$ masses by those of the $K^-$ and $\eta$ mesons.
A more detailed derivation of the differential distribution in the $K\eta$ case
can be found in ref.~\cite{Escribano:2013bca}. Regarding the global
normalisation, in the following we will employ
$|V_{us}f_+^{K_S\pi^-}(0)|=0.2163(5)$ \cite{Antonelli:2010yf}, from a global
fit to $K_{\ell3}$ data, and
$|V_{us}f_+^{K^-\eta}(0)|=|V_{us}f_+^{K_S\pi^-}(0)|\cos\theta_P$, with
$\theta_P=-(13.3\pm1.0)^\circ$ \cite{Ambrosino:2006gk}.

The required form factors cannot be computed analytically from first principles.
Still, the symmetries of the underlying QCD Lagrangian are useful to determine
their behaviour in specific limits, the chiral or low-energy limit and the
high-energy behaviour, so that the model dependence is reduced to the
interpolation between these known regimes. For our central fits, to be presented
in the next section, we follow the dispersive representation of the vector form
factors outlined in ref.~\cite{Boito:2008fq}, and briefly summarised below for
the convenience of the reader.  For the case of the $K_S\pi^-$ system, including
two resonances, the $K^*=K^*(892)$ and the $K^{*\prime}=K^*(1410)$, the reduced
vector form factor is taken to be of the form \cite{Boito:2008fq}
\begin{equation}
\label{FpKpi2}
\widetilde{f}_+^{K\pi}(s) \,=\, \frac{m_{K^*}^2 - \kappa_{K^*}\,
\widetilde{H}_{K\pi}(0) + \gamma s}{D(m_{K^*},\gamma_{K^*})} -
\frac{\gamma s}{D(m_{K^{*\prime}},\gamma_{K^{*\prime}})} \,,
\end{equation}
where
\begin{equation}
\label{Dden}
D(m_n,\gamma_n) \,=\, m_n^2 - s - \kappa_n \widetilde{H}_{K\pi}(s) \,,
\end{equation}
and
\begin{equation}
\label{kappa}
\kappa_n \,=\, \frac{192\pi}{\sigma_{K\pi}(m_n^2)^3}\frac{\gamma_n}{m_n} \,.
\end{equation}
The fit function for the vector form factor is expressed in terms of the
unphysical ``mass'' and ``width'' parameters $m_n$ and $\gamma_n$. They are
denoted by small letters, to distinguish them from the physical mass and width
parameters $M_n$ and $\Gamma_n$, which will later be determined from the pole
positions in the complex plane and are denoted by capital letters. The scalar
one-loop integral function $\widetilde{H}_{K\pi}(s)$ is defined below eq.~(3) of
ref.~\cite{Jamin:2006tk}, however removing the factor $1/f_\pi^2$ which cancels
if $\kappa_n$ is expressed in terms of the unphysical width $\gamma_n$. Finally,
in eq.~(\ref{kappa}), the phase space function $\sigma_{K\pi}(s)$ is given by
$\sigma_{K\pi}(s)=2q_{K\pi}(s)/\sqrt{s}$. Since the $K^*$ resonances that are
produced through the $\tau$ decay are charged, and can decay or rescatter into
both $K^0\pi^-$ as well as $K^-\pi^0$ channels, in the resonance propagators
described by eqs.~(\ref{FpKpi2}) to (\ref{kappa}) we have chosen to employ the
corresponding isospin average, that is
\begin{equation}
\label{HtKpi}
\widetilde{H}_{K\pi}(s) \,=\, \frac{2}{3}\,\widetilde{H}_{K^0\pi^-}(s) +
                              \frac{1}{3}\,\widetilde{H}_{K^-\pi^0}(s) \,,
\end{equation}
and analogously for $\sigma_{K\pi}(s)$, such that the resonance width contains
both contributions.
Little is known about a proper description of the width of the second vector
resonance $K^{*\prime}$. The complicated $K^*\pi \sim K\pi\pi$ cuts may yield
relevant effects which however necessitates a coupled-channel analysis like in
refs.~\cite{Moussallam:2007qc,Bernard:2013jxa}. This is beyond the scope of
the present paper, in which for simplicity also for the second resonance only
the two-meson cut is included. Similar remarks apply to a proper inclusion of
the $K\eta$ and $K\eta'$ channels into eq.~(\ref{HtKpi}) which would also
require a coupled-channel analysis as was done for the corresponding scalar
form factors in refs.~\cite{JOP00,JOP02}.

Next, we further follow ref.~\cite{Boito:2008fq} in writing a three-times
subtracted dispersive representation for the vector form factor,
\begin{equation}
\label{dispersive VFF}
\widetilde{f}_+^{K\pi}(s) \,=\, \exp\Biggl[\, \alpha_1\frac{s}{M_{\pi^-}^2} +
\frac{1}{2}\alpha_2\frac{s^2}{M_{\pi^-}^4} +\frac{s^3}{\pi}
\int\limits_{s_{K\pi}}^{s_{\rm cut}} {\rm d}s'
\frac{\delta_1^{K\pi}(s')}{(s')^3(s'-s-i0)} \,\Biggr] \,,
\end{equation}
where $s_{K\pi}=(M_K+M_\pi)^2$ is the $K\pi$ threshold\footnote{Isospin breaking
on the low-energy parameters, like the threshold of the dispersive integral or
the slope parameters of the vector form factor, is discussed later on.} and the
two subtraction constants $\alpha_1$ and $\alpha_2$ are related to the slope
parameters appearing in the low-energy expansion of the form factor:
\begin{equation}
\label{slope parameters}
\widetilde{f}_+^{K\pi}(s) \,=\, 1 + \lambda_+^{'}\frac{s}{M_{\pi^-}^2} +
\frac{1}{2}\lambda_+^{''}\frac{s^2}{M_{\pi^-}^4} +
\frac{1}{6}\lambda_+^{'''}\frac{s^3}{M_{\pi^-}^6} + \ldots \,.
\end{equation}
Explicitly, the relations for the linear and quadratic slope parameters
$\lambda_+^{'}$ and $\lambda_+^{''}$ take the form:
\begin{equation}
\label{lambdaalpha}
\lambda_+^{'}  \,=\, \alpha_1 \,, \qquad
\lambda_+^{''} \,=\, \alpha_2 + \alpha_1^2 \,.
\end{equation}
The incentive for employing a dispersive representation for the form factor
is that in this way the influence of the less-well known higher energy region
is suppressed. The associated error can be estimated by varying the cut-off
$s_{\rm cut}$ in the dispersive integral. In order to obtain the required input
phase $\delta_1^{K\pi}(s)$, like in \cite{Boito:2008fq} we use the resonance
propagator representation eq.~(\ref{FpKpi2}) of the vector form factor.
The phase can then be calculated from the relation
\begin{equation}
\label{del1Kpi2}
\tan\delta_1^{K\pi}(s) = \frac{{\rm Im}\widetilde{f}_+^{K\pi}(s)}
                              {{\rm Re}\widetilde{f}_+^{K\pi}(s)} \,,
\end{equation}
which completes our representation of the vector form factor
$\widetilde{f}_+^{K\pi}(s)$.

The scalar form factors that are required for a complete description of the
decay spectra according to eq.~(\ref{spectral function}) will be taken from
the coupled-channel dispersive representation of refs.~\cite{JOP00,JOP02}. In
particular, for the scalar $K\pi$ form factor, we employ the update presented
in ref.~\cite{JOP06}. For the scalar $K\eta$ form factor, the result of the
three-channel analysis described in section~4.3 of \cite{JOP02} is used,
choosing specifically the solution corresponding to fit~(6.10) of
ref.~\cite{JOP00}. As a matter of principle, this is not fully consistent,
since the employed $K\pi$ form factor was extracted from a two-channel analysis,
only including the dominant $K\pi$ and $K\eta'$ channels. But as our numerical
analysis shows, anyway the influence of the scalar $K\eta$ form factor is
insignificant so that this inconsistency can be tolerated.

\section{Joint fits to
\texorpdfstring{\boldmath{$\tau^- \to K_S\pi^-\nu_\tau$}}{tau- to KS pi- nutau}
and
\texorpdfstring{\boldmath{$\tau^- \to K^-\eta\nu_\tau$}}{tau- to K- eta nutau}
Belle
data}\label{Fits}

The differential decay rate of eq.~(\ref{spectral function}) is related to the
distribution of the measured number of events by means of
\begin{equation}
\label{theory_to_experiment}
\frac{{\rm d}N_{\rm events}}{{\rm d}\sqrt{s}} \,=\,
\frac{{\rm d}\Gamma(\tau^-\to (PQ)^-\nu_\tau)}{{\rm d}\sqrt{s}}\,
\frac{N_{\rm events}}{\Gamma_\tau \bar{B}(\tau^-\to (PQ)^-\nu_\tau)}
\,\Delta \sqrt{s_{\rm bin}} \,,
\end{equation}
where $N_{\rm events}$ is the total number of events measured for the
considered process, $\Gamma_\tau$ is the inverse $\tau$ lifetime and
$\Delta \sqrt{s_{\rm bin}}$ is the bin width. 
$\bar{B}(\tau^-\to (PQ)^-\nu_\tau)\equiv \bar{B}_{PQ}$ is a normalisation
constant that, for a perfect description of the spectrum, would equal the
corresponding branching fraction.

For the $\tau^-\to K_S\pi^-\nu_\tau$ decays, an unfolded distribution measured
by Belle is available \cite{Epifanov:2007rf}. The corresponding number of
events is $53113.21$ ($54157.59$ before unfolding) and the bin width
$11.5\,$MeV. As discussed in the earlier analyses, the data points corresponding
to bins $5$, $6$ and $7$ are difficult to bring into accord with the theoretical
descriptions and have thus been excluded from the minimisation.\footnote{Still,
including them in the fits would just increase the $\chi^2$ with only
irrelevant changes in the fit parameters.} The first point has not been
included either, since the centre of the bin lies below the $K_S\pi^-$
production threshold. Following a suggestion from the experimentalists, as in
the previous analyses we have furthermore excluded data corresponding to bin
numbers larger than $90$.

On the other hand, the published $\tau^-\to K^-\eta\nu_\tau$ Belle data
\cite{Inami:2008ar} are only available still folded with detector
effects.\footnote{Contrary to our previous analysis \cite{Escribano:2013bca}, 
in the present study we have not included the BaBar data
\cite{delAmoSanchez:2010pc}. They only consist in ten data points, with rather
large errors, which furthermore had to be digitised from the published plots.}
Lacking for a better alternative, we have assumed that the $K^-\eta$ unfolding
function is reasonably estimated by the $K_S\pi^-$ one and we have extracted 
in this way pseudo-unfolded data that we employed in our analysis. The
corresponding number of events turns out $1271.51$ for a bin width of $25$ MeV.
In this case, we excluded the first three data points, which lie below the
$K^-\eta$ production threshold, and discarded data above the $\tau$ mass.

The $\chi^2$ function minimised in our fits was chosen to be
\begin{equation}
\label{chi^2}
\chi^2 \,=\, \sum_{i,\;PQ=K_S\pi^-,\,K^-\eta}{}^{\hspace{-10.5mm}{}^\prime}
\hspace{9mm}
\left(\frac{\mathcal{N}_i^{th}-\mathcal{N}_i^{exp}}{\sigma_{\mathcal{N}_i^{exp}}}\right)^2 +
\sum_{PQ=K_S\pi^-,\,K^-\eta}\left(\frac{\bar{B}_{PQ}^{th}-B_{PQ}^{exp}}{\sigma_{B_{PQ}}^{exp}}\right)^2 \,,
\end{equation}
where $\mathcal{N}_i^{exp}$ and $\sigma_{\mathcal{N}_i^{exp}}$ are,
respectively, the experimental number of events and the corresponding
uncertainties in the $i$-th bin.\footnote{While it is expected that bin-to-bin
correlations due to unfolding should arise, a full covariance matrix for the
spectral data is not available, whence we have to limit ourselves to the
diagonal errors.} The prime in the summation indicates that the points specified
above have been excluded. Therefore, the number of fitted data points is $86$
($28$) for the $K_S\pi^-$ ($K^-\eta$) spectrum, together with the respective
branching fractions: hence $116$ data points in total. While it is possible to
obtain stable fits without using the $K_S\pi^-$ branching fraction as a data
point, this is not the case for the $K^-\eta$ channel. This is due to the fact
that there are strong correlations between the branching ratio and the slope
parameters of the vector form factor. While in the $K_S\pi^-$ case sufficiently
many data points with small enough errors are available to determine all fit
quantities from the spectrum, for the $K^-\eta$ decay mode this was not
possible. As a consistency check, we will be comparing the fitted values of the
respective branching ratios to the corresponding results obtained by directly
integrating the spectrum in all our fits.

The fitted parameters within the dispersive representation of the form factors
of eq.~(\ref{dispersive VFF}) then include:
\begin{itemize}
\item the respective branching fractions $\bar{B}_{K\pi}$ and $\bar{B}_{K\eta}$.
For consistency, as our inputs in eq.~(\ref{chi^2}) we employ the results
obtained by Belle in correspondence with the employed decay distribution data:
$(0.404 \pm 0.013)\%$ \cite{Epifanov:2007rf} as well as
$(1.58\pm0.10)\times10^{-4}$ \cite{Inami:2008ar}, respectively. This may be
compared to the averages by the Particle Data Group, $(0.420 \pm 0.020)\%$ and
$(1.52 \pm 0.08)\times10^{-4}$ \cite{Beringer:1900zz} and Heavy Flavour Averaging
Group values \cite{Amhis:2012bh}, $(0.410 \pm 0.009)\%$ and
$(1.53 \pm 0.08)\times10^{-4}$. The recent update by Belle \cite{Ryu:2014vpc}
including a 669 fb$^{-1}$ data sample was found to be $(0.416 \pm 0.008)\%$
for the former decay mode.

\item The slope parameters: $\lambda^{\prime(\prime)}_{K\pi}$ and
$\lambda^{\prime(\prime)}_{K\eta}$. As was noted in
ref.~\cite{Escribano:2013bca}, while the former ones correspond to the
$K_S\pi^-$ channel, the latter ones are related to the $K^-\pi^0$ system.
Therefore, small differences in these parameters due to isospin violations are
expected, and in the most general fit we allow for independent parameters in
the two channels. As consistency checks of our procedure, we have also
considered some fits assuming $\lambda^\prime_{K\eta}=\lambda^\prime_{K\pi}$.
The findings of ref.~\cite{Boito:2008fq}, 
$\lambda^\prime_{K\pi}=(24.66 \pm 0.77)\times10^{-3}$ and
$\lambda^{\prime\prime}_{K\pi}=(11.99 \pm 0.20)\times10^{-4}$, should serve as
a reference point for our present study, where however $\bar{B}_{K\pi}$ was
fixed to the average $(0.418 \pm 0.011)\%$ at that time.

\item The pole parameters of the $K^*(892)$ and $K^*(1410)$ resonances. The
masses and widths of these resonances are extracted from the complex pole
position $s_R$ according to $\sqrt{s_R}=M_R-\frac{i}{2}\Gamma_R$
\cite{Escribano:2002iv}. For the lowest-lying resonance our results for the
pole mass and width should be compatible with $(892.0 \pm 0.2)\,$MeV and
$(46.2 \pm 0.4)\,$MeV \cite{Boito:2010me}, respectively, where the quoted
uncertainties are only statistical. We expect that the extraction of the
$K^*(1410)$ pole position should benefit from our present combined fit for
which $(1273 \pm 75)\,$MeV and $(185 \pm 74)\,$MeV were obtained in
ref.~\cite{Boito:2008fq} when the uncertainties are symmetrised.

\item The relative weight $\gamma$ of the two resonances. In our
isospin-symmetric way (\ref{FpKpi2}) of parametrising the resonance
propagators in the form factor description, $\gamma$ should be the same for
the $K_S\pi^-$ and $K^-\eta$ channels, which we shall assume for our central
fit. Still, we have also tried to fit them independently, as differences
might indicate inelastic or coupled-channel effects. As is seen below, our
various fit results do not show a sizeable preference for this possibility
which supports our choice $\gamma_{K\eta}=\gamma_{K\pi}$. 
Our findings may be compared to the value  $\gamma=-0.039\pm0.020$ of
\cite{Boito:2008fq} indicating the influence of including the
$\tau^-\to K^-\eta\nu_\tau$ mode into our analysis.
\end{itemize}
In the fits we have furthermore employed the following numerical inputs:
$M_\tau=1776.82\,$MeV, $\Gamma_\tau=2.265\times10^{-12}\,$GeV and
$G_F=1.16637(1)\times10^{-5}\,$GeV$^{-2}$ \cite{Beringer:1900zz}.
Pseudoscalar meson masses were also taken according to their PDG values
\cite{Beringer:1900zz}. Finally, the next-to-leading order $\chi PT$ low-energy
constants and the chiral logarithms depend on an arbitrary renormalisation
scale $\mu$ (these dependencies cancel one another), which we have fixed to
the physical mass scale of the problem, $M_{K^*}=892\,$MeV.

In Table~\ref{Tab:DifferentFits}, we display our results using slightly
different settings, though in all of them eq.~(\ref{del1Kpi2}) is employed to
obtain the input phaseshift for the dispersion relation (\ref{dispersive VFF})
and $s_{\rm cut}$ is fixed to $4\,$GeV$^2$ (the uncertainty associated to its
variation is discussed later on): our reference fit (second column) 
corresponds to fixing $\gamma_{K\pi}=\gamma_{K\eta}$, fit~A (third column)
assumes $\lambda'_{K\pi}=\lambda'_{K\eta}$, fit~B (fourth column) is the
result of letting all parameters float independently and finally, fit~C
(fifth column) enforces both restrictions $\gamma_{K\pi}=\gamma_{K\eta}$ and
$\lambda^{\prime}_{K\pi}=\lambda^{\prime}_{K\eta}$. It is seen that our
approach is rather stable against these variations, as the $\chi^2/$n.d.f.
remains basically the same for the different scenarios. Also the values of 
the fitted parameters are always compatible across all fits. The largest
modification is observed in fit~A, where we fix
$\lambda'_{K\pi}=\lambda'_{K\eta}$, but allow for independent resonance mixing
parameters $\gamma$. This is partly expected since in the reference fit the
former equality on the slope parameters is only fulfilled at the $2\sigma$
level. Letting all parameters float in fit~B yields results which are nicely
compatible with the reference fit, though for some parameters resulting in
slightly larger uncertainties. Finally, enforcing both, the linear slopes as
well as the mixing parameters to be equal also results in a compatible fit
where now the largest shift by about $2\sigma$ is found in
$\lambda^{''}_{K\eta}$.

\begin{table}
\begin{center}
\begin{tabular}{|c|c|c|c|c|}
\hline
Fitted value& Reference Fit & Fit A & Fit B & Fit C\cr
\hline
$\bar{B}_{K\pi}(\%)$ & $0.404\pm0.012$ & $0.400\pm0.012$ & $0.404\pm0.012$ & $0.397\pm0.012$\cr
$(B_{K\pi}^{th})(\%)$ & $(0.402)$ & $(0.394)$ & $(0.400)$ & $(0.394)$\cr
$M_{K^*}$ & $892.03\pm0.19$ & $892.04\pm0.19$ & $892.03\pm0.19$ & $892.07\pm0.19$\cr
$\Gamma_{K^*}$ & $46.18\pm0.42$ & $46.11\pm0.42$ & $46.15\pm0.42$ & $46.13\pm0.42$\cr
$M_{K^{*\prime}}$ & $1305^{+15}_{-18}$ & $1308^{+16}_{-19}$ & $1305^{+15}_{-18}$ & $1310^{+14}_{-17}$\cr
$\Gamma_{K^{*\prime}}$ & $168^{+52}_{-44}$ &$212^{+66}_{-54}$ & $174^{+58}_{-47}$ & $184^{+56}_{-46}$\cr
$\gamma_{K\pi}\times10^2$ & $=\gamma_{K\eta}$ & $-3.6^{+1.1}_{-1.5}$ & $-3.3^{+1.0}_{-1.3}$ & $=\gamma_{K\eta}$\cr
$\lambda^{\prime}_{K\pi}\times10^3$ & $23.9\pm0.7$ & $23.6\pm0.7$ & $23.8\pm0.7$ & $23.6\pm0.7$\cr
$\lambda^{\prime\prime}_{K\pi}\times10^4$ & $11.8\pm0.2$ & $11.7\pm0.2$ & $11.7\pm0.2$ & $11.6\pm0.2$\cr
$\bar{B}_{K\eta}\times10^4$ & $1.58\pm0.10$ & $1.62\pm0.10$ & $1.57\pm0.10$ & $1.66\pm0.09$\cr
$(B_{K\eta}^{th})\times10^4$ & $(1.45)$ & $(1.51)$ & $(1.44)$ & $(1.58)$\cr
$\gamma_{K\eta}\times10^2$ & $-3.4^{+1.0}_{-1.3}$ & $-5.4^{+1.8}_{-2.6}$ & $-3.9^{+1.4}_{-2.1}$ & $-3.7^{+1.0}_{-1.4}$\cr
$\lambda^{\prime}_{K\eta}\times10^3$ & $20.9\pm1.5$ & $=\lambda^{\prime}_{K\pi}$ & $21.2\pm1.7$ & $=\lambda^{\prime}_{K\pi}$\cr
$\lambda^{\prime\prime}_{K\eta}\times10^4$ & $11.1\pm0.4$ & $11.7\pm0.2$ & $11.1\pm0.4$ & $11.8\pm0.2$\cr
$\chi^2/$n.d.f. & $108.1/105\sim1.03$ & $109.9/105\sim1.05$ & $107.8/104\sim1.04$ & $111.9/106\sim1.06$ \cr
\hline
\end{tabular}
\caption{\label{Tab:DifferentFits} \small{Fit results for different choices
regarding linear slopes and resonance mixing parameters at
$s_{\rm cut}=4\,$GeV$^2$. See the main text for further details. Dimensionful
parameters are given in MeV. As a consistency check, for each of the fits we
provide (in brackets) the value of the respective branching fractions obtained
by integrating eq.~(\ref{spectral function}).}}
\end{center}
\end{table}

The theoretical uncertainty associated to the choice of $s_{\rm cut}$ is probed
through the fits presented in Table~\ref{Tab:ReferenceFit} where, for the
setting of our reference fit discussed previously, the values $3.24\,$GeV$^2$
(second column), $4\,$GeV$^2$ (third column), $9\,$GeV$^2$ (fourth column) and
the $s_{\rm cut}\to\infty$ limit (last column) are used
($s_{\rm cut}=4\,$GeV$^2$ corresponds to our reference fit in the second column
of Table \ref{Tab:DifferentFits} and is repeated here for ease of comparison).
The dependence of the fitted parameters on the integral cut-off is similar to
what was found in previous works (see, for instance
refs.~\cite{Boito:2008fq, Boito:2010me}) and allows to estimate the 
corresponding systematic error. In order to corroborate our fits, we performed
additional tests. We have also run fits considering two and four subtraction
constants in order to test the stability of our results with respect to this
choice. As in the previous analyses \cite{Boito:2008fq,Boito:2010me} of the
$\tau^- \to K_S\pi^-\nu_\tau$ spectrum, the changes in the results are well
within our uncertainties. It is furthermore confirmed that regarding final
uncertainties three subtractions appears to be an optimal choice. This may,
however, change if the representation of the higher-energy region is improved,
for example through a coupled-channel analysis, such that this region requires
less suppression. As a second test, we have employed a variant of the form
factor Ansatz (\ref{FpKpi2}) in which the real part of the loop function
$\widetilde{H}_{K\pi}(s)$ is not resummed into the propagator denominator,
but into an exponential, as was for example suggested in
refs.~\cite{Jamin:2006tk,Jamin:2008qg} for the description of
$\tau \to K\pi\nu_\tau$ decays. This type of Ansatz is further discussed in
Appendix~A where also direct fits of the corresponding form factors are
described. Our test here, however, consists in extracting the corresponding
phase from this type of form factor according to eq.~(\ref{del1Kpi2}) and
plugging the respective phase into the dispersion relation
(\ref{dispersive VFF}). It is found that the corresponding fits are almost
identical to the ones described before, providing additional faith on the
robustness of the extracted parameters.

\begin{table}
\begin{center}
\begin{tabular}{|c|c|c|c|c|}
\hline
\backslashbox{Fitted value}{$s_{\rm cut}$(GeV$^2$)}& $3.24$ & $4$ & $9$ &
$\infty$\cr
\hline
$\bar{B}_{K\pi}(\%)$ & $0.402\pm0.013$ & $0.404\pm0.012$ & $0.405\pm0.012$ & $0.405\pm0.012$\cr
$(B_{K\pi}^{th})(\%)$ & $(0.399)$ & $(0.402)$ & $(0.403)$ & $(0.403)$\cr
$M_{K^*}$ & $892.01\pm0.19$ & $892.03\pm0.19$ & $892.05\pm0.19$ & $892.05\pm0.19$\cr
$\Gamma_{K^*}$ & $46.04\pm0.43$ & $46.18\pm0.42$ & $46.27\pm0.42$ & $46.27\pm0.41$\cr
$M_{K^{*\prime}}$ & $1301^{+17}_{-22}$ & $1305^{+15}_{-18}$ & $1306^{+14}_{-17}$ & $1306^{+14}_{-17}$\cr
$\Gamma_{K^{*\prime}}$ & $207^{+73}_{-58}$ &$168^{+52}_{-44}$ & $155^{+48}_{-41}$ & $155^{+47}_{-40}$\cr
$\gamma_{K\pi}$ & $=\gamma_{K\eta}$ & $=\gamma_{K\eta}$ & $=\gamma_{K\eta}$ & $=\gamma_{K\eta}$\cr
$\lambda^{\prime}_{K\pi}\times10^3$ & $23.3\pm0.8$ & $23.9\pm0.7$ & $24.3\pm0.7$ & $24.3\pm0.7$\cr
$\lambda^{\prime\prime}_{K\pi}\times10^4$ & $11.8\pm0.2$ & $11.8\pm0.2$ & $11.7\pm0.2$ & $11.7\pm0.2$\cr
$\bar{B}_{K\eta}\times10^4$ & $1.57\pm0.10$ & $1.58\pm0.10$ & $1.58\pm0.10$ & $1.58\pm0.10$\cr
$(B_{K\eta}^{th})\times10^4$ & $(1.43)$ & $(1.45)$ & $(1.46)$ & $(1.46)$\cr
$\gamma_{K\eta}\times10^2$ & $-4.0^{+1.3}_{-1.9}$ & $-3.4^{+1.0}_{-1.3}$ & $-3.2^{+0.9}_{-1.1}$ & $-3.2^{+0.9}_{-1.1}$\cr
$\lambda^{\prime}_{K\eta}\times10^3$ & $18.6\pm1.7$ & $20.9\pm1.5$ & $22.1\pm1.4$ & $22.1\pm1.4$\cr
$\lambda^{\prime\prime}_{K\eta}\times10^4$ & $10.8\pm0.3$ & $11.1\pm0.4$ & $11.2\pm0.4$ & $11.2\pm0.4$\cr
$\chi^2/$n.d.f. & $105.8/105$ & $108.1/105$ & $111.0/105$ & $111.1/105$ \cr
\hline
\end{tabular}
\caption{\label{Tab:ReferenceFit} \small{Reference fit results obtained for
different values of $s_{\rm cut}$ in the dispersive integral are displayed.
Dimensionful parameters are given in MeV. As a consistency check, for each of
the fits we give (in brackets) the value of the respective branching ratios
obtained integrating eq.~(\ref{spectral function}).}}
\end{center}
\end{table}

\begin{figure}[thb]
\begin{center}
\vspace*{1.25cm}
\includegraphics[scale=0.85]{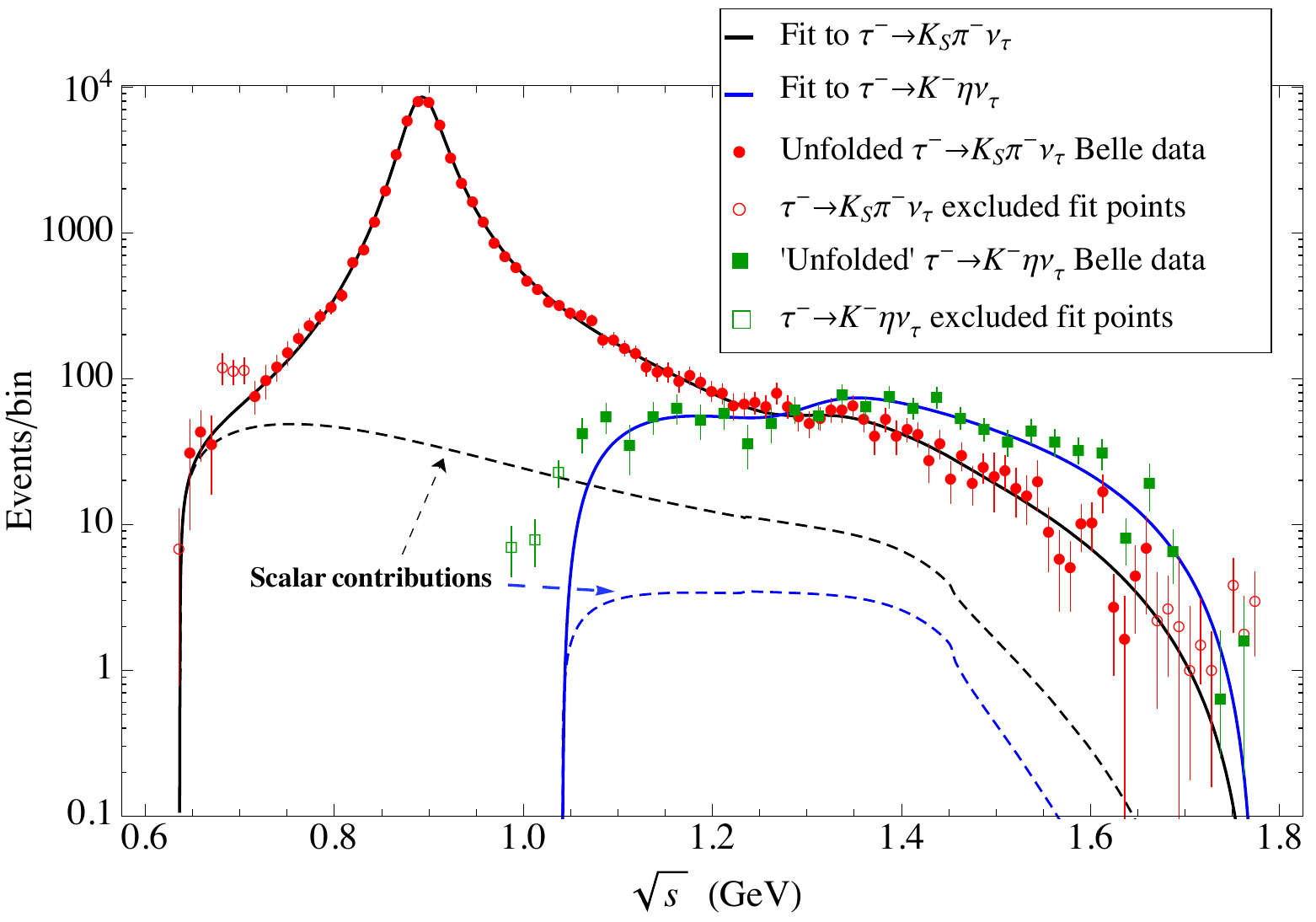}
\caption{\label{fig:Spectrum} \small{Belle $\tau^-\to K_S\pi^-\nu_\tau$
(red solid circles) \cite{Epifanov:2007rf} and $\tau^-\to K^-\eta\nu_\tau$
(green solid squares) \cite{Inami:2008ar} measurements as compared to our best
fit results (solid black and blue lines, respectively) obtained in combined
fits to both data sets, as presented in eq.~(\ref{Final Results}). Empty
circles (squares) correspond to data points which have not been included in
the analysis. The small scalar contributions have been represented by black
and blue dashed lines showing that while the former plays a role for the $K\pi$
spectrum close to threshold, the latter is irrelevant for the $K\eta$
distribution.}}
\end{center}
\end{figure}

For presenting our final results, we have added to the statistical fit error
a systematic uncertainty due to the variation of $s_{\rm cut}$. To this end, we
have taken the largest variation of central values while varying $s_{\rm cut}$
(which is always found at $s_{\rm cut}=3.24\,$GeV$^2$) and have added this
variation in quadrature to the statistical uncertainty. We then obtain
\begin{eqnarray}
\label{Final Results}
& & \bar{B}_{K\pi}\,=\,\left(0.404\pm0.012\right)\%\,,\quad M_{K^*}\,=\, 892.03\pm0.19\,,\quad \Gamma_{K^*}\,=\,46.18\pm0.44\,,\nonumber\\[2mm]
& & M_{K^{*\prime}}\,=\, 1305^{+16}_{-18}\,,\quad \Gamma_{K^{*\prime}}\,=\,168^{+65}_{-59}\,,\quad \gamma_{K\pi}=\gamma_{K\eta}=\left(-3.4^{+1.2}_{-1.4}\right)\cdot10^{-2}\,,\nonumber\\[2mm]
& & \lambda^{\prime}_{K\pi}\,=\,\left(23.9\pm0.9\right)\cdot10^{-3}\,,\quad \lambda^{\prime\prime}_{K\pi}\,=\,\left(11.8\pm0.2\right)\cdot10^{-4}\,,\quad \bar{B}_{K\eta}\,=\,\left(1.58\pm0.10\right)\cdot10^{-4}\,,\nonumber\\[2mm]
& & \lambda^{\prime}_{K\eta}\,=\,\left(20.9\pm2.7\right)\cdot10^{-3}\,,\quad \lambda^{\prime\prime}_{K\eta}\,=\,\left(11.1\pm0.5\right)\cdot10^{-4}\,,
\end{eqnarray}
were like before all dimensionful quantities are given in MeV. Our final fit
results are compared to the measured Belle $\tau^-\to K_S\pi^-\nu_\tau$ and
$\tau^-\to K^-\eta\nu_\tau$ distributions \cite{Epifanov:2007rf,Inami:2008ar}
in Figure~\ref{fig:Spectrum}. Satisfactory agreement with the experimental data,
in accord with the observed $\chi^2$/n.d.f.~of order one, is seen for all data
points. The $K\pi$ spectrum is dominated by the contribution of the $K^*(892)$
resonance, whose peak is neatly visible. The scalar form factor contribution,
although small in most of the phase space, is important to describe the data
immediately above threshold. There is no such clear peak structure for the
$K\eta$ channel as a consequence of the interplay between both $K^*$ resonances.
The corresponding scalar form factor in this case is numerically insignificant.

The correlation coefficients corresponding to our reference fit with
$s_{\rm cut}=4$ GeV$^2$ can be read from Table \ref{Tab:CorrMat}. As
anticipated, there is a large correlation between the set
$\{\bar{B}_{K\pi},\,\lambda'_{K\pi},\,\lambda^{''}_{K\pi}\}$ which enables
stable fits removing one of these parameters (the fit then becomes somewhat
less restrictive, though). Despite the correlation between $\lambda'_{K\eta}$
and $\lambda^{''}_{K\eta}$ also being nearly maximal, these parameters are less
correlated with $\bar{B}_{K\eta}$, implying that all three are needed to reach
convergence in the minimisation. For this reason we prefer to keep
$\bar{B}_{K\eta}$ as a data point in the joint analysis. Finally, we note a
large correlation between the parameters $\gamma_{K\pi}=\gamma_{K\eta}$ and
$\Gamma_{K^{*\prime}}$ which seems to be enhancing the corresponding errors
(this effect may in part be due to the three subtractions employed, which
decrease the sensitivity to the higher-energy region). In the fits where
$\gamma_{K\pi}=\gamma_{K\eta}$ is not enforced, their correlation coefficient
is $\approx 0.67$. This suggests that with more precise data in the future it
might be possible to resolve the current degeneracy between both.

\begin{table}
\begin{center}
%\rotatebox{90}{
\begin{tabular}{|c|c|c|c|c|c|c|c|c|c|c|c|}
\hline
 & \tiny{$\bar{B}_{K\pi}$} & \tiny{$M_{K^*}$} & \tiny{$\Gamma_{K^*}$} & \tiny{$M_{K^{*\prime}}$} & \tiny{$\Gamma_{K^{*\prime}}$} & \tiny{$\lambda^{\prime}_{K\pi}$} & \tiny{$\lambda^{\prime\prime}_{K\pi}$} & \tiny{$\bar{B}_{K\eta}$} & \tiny{$\gamma_{K\eta}=\gamma_{K\pi}$} & \tiny{$\lambda^{\prime}_{K\eta}$} & \tiny{$\lambda^{\prime\prime}_{K\eta}$} \cr
\hline
\tiny{$M_{K^*}$} & \tiny{$-0.163$} & \tiny{$1$} &  &  &  &  &  &  &  &  & \cr
\tiny{$\Gamma_{K^*}$} & \tiny{$0.028$} & \tiny{$-0.060$} & \tiny{$1$} &  &  &  &  &  &  &  & \cr
\tiny{$M_{K^{*\prime}}$} & \tiny{$-0.063$} & \tiny{$-0.104$} & \tiny{$-0.142$} & \tiny{$1$} &  &  &  &  &  & & \cr
\tiny{$\Gamma_{K^{*\prime}}$} & \tiny{$0.126$} & \tiny{$0.130$} & \tiny{$0.292$} & \tiny{$-0.556$} & \tiny{$1$} &  &  &  &  &  & \cr
\tiny{$\lambda^{\prime}_{K\pi}$} & \tiny{$0.800$} & \tiny{$-0.100$} & \tiny{$0.457$} & \tiny{$-0.244$} & \tiny{$0.432$} & \tiny{$1$} &  &  &  &  & \cr
\tiny{$\lambda^{\prime\prime}_{K\pi}$} & \tiny{$0.928$} & \tiny{$-0.215$} & \tiny{$0.328$} & \tiny{$-0.166$} & \tiny{$0.304$} & \tiny{$0.942$} & \tiny{$1$} &  &  &  & \cr
\tiny{$\bar{B}_{K\eta}$} & \tiny{$-0.003$} & \tiny{$-0.005$} & \tiny{$-0.010$} & \tiny{$0.003$} & \tiny{$-0.001$} & \tiny{$-0.015$} & \tiny{$-0.009$} & \tiny{$1$} &  &  & \cr
\tiny{$\gamma_{K\eta}=\gamma_{K\pi}$} & \tiny{$-0.155$} & \tiny{$-0.173$} & \tiny{$-0.378$} & \tiny{$0.498$} & \tiny{$-0.878$} & \tiny{$-0.565$} & \tiny{$-0.373$} & \tiny{$0.019$} & \tiny{$1$} &  & \cr
\tiny{$\lambda^{\prime}_{K\eta}$} & \tiny{$0.058$} & \tiny{$0.028$} & \tiny{$0.117$} & \tiny{$0.050$} & \tiny{$0.337$} & \tiny{$0.182$} & \tiny{$0.128$} & \tiny{$0.434$} & \tiny{$-0.340$} & \tiny{$1$} & \cr
\tiny{$\lambda^{\prime\prime}_{K\eta}$} & \tiny{$0.035$} & \tiny{$-0.017$} & \tiny{$0.037$} & \tiny{$0.106$} & \tiny{$0.218$} & \tiny{$0.080$} & \tiny{$0.064$} & \tiny{$0.561$} & \tiny{$-0.174$} & \tiny{$0.971$} & \tiny{$1$} \cr
\hline
\end{tabular}
%}
\caption{\label{Tab:CorrMat} \small{Correlation coefficients corresponding
to our reference fit with $s_{\rm cut}=4$ GeV$^2$, second column of
Table~\ref{Tab:DifferentFits}. In the fits where $\gamma_{K\pi}=\gamma_{K\eta}$
is not enforced, their correlation coefficient turns out to be $\approx 0.67$.}}
\end{center}
\end{table}

Several comments regarding our final results of eq.~(\ref{Final Results})
and the reference fit of Table~\ref{Tab:DifferentFits} are in order:
\begin{itemize}
\item Concerning the branching fractions, we observe that in the $K_S\pi^-$
channel our fit value $\bar{B}_{K\pi}$, which is mainly driven by the explicit
input, and the result when integrating the fitted spectrum $B_{K\pi}^{th}$,
are in very good agreement, pointing to a satisfactory description of the
experimental data. On the other hand, for the $K\eta$ case, one notes a trend
that the integrated branching fraction $B_{K\eta}^{th}$ turns out about $10\%$
smaller than the fit result $\bar{B}_{K\eta}$, which points to slight
deficiencies in the theoretical representation of this spectrum. This issue
should be investigated further in the future with more precise data.

\item The $K_S\pi^-$ slope parameters are well compatible with previous
analogous analysis \cite{Boito:2008fq,Boito:2010me}. For the corresponding
$K^-\eta$ slopes, we obtain somewhat smaller values, which are, however,
compatible with the crude estimates in Ref.~\cite{Escribano:2013bca}. The fact
that the $K^-\eta$ slopes are about $2\sigma$ lower than the $K_S\pi^-$
slopes could be an indication of isospin violations, or could be a purely
statistical effect. (Or a mixture of both.) To tackle this question and make
further progress to disentangle isospin violations in the $K\pi$ form factor
slopes, it is indispensable to study the related distribution for the
$\tau^-\to K^-\pi^0\nu_\tau$ decay, and the experimental groups should make
every effort to also publish the corresponding spectrum for this process.

\item The pole parameters of the $K^*(892)$ resonance are in nice accord with
previous values \cite{Boito:2008fq,Boito:2010me} and have similar statistical
fit uncertainties which is to be expected as these parameters are driven by the
data of the $\tau^-\to K_S\pi^-\nu_\tau$ decay, which was the process analysed
previously. Regarding the parameters of the $K^*(1410)$ resonance, adding the
$\tau^-\to K^-\eta\nu_\tau$ spectral data into the fit results in a substantial
improvement in the determination of the mass, while only a slight improvement
in the width is observed. Part of the large uncertainty in the width of the
second $K^*$ resonance can be traced back to the strong fit correlation with
the mixing parameter $\gamma$, which is also not very well determined.
Future data of either $\tau^-\to(K\pi)^-\nu_\tau$ or $\tau^-\to K^-\eta\nu_\tau$
hadronic invariant mass distributions should enable a more precise evaluation.
Prospects updating the Belle-I analyses with the complete data sample or
studying Belle-II data are discussed next.
\end{itemize}

In Table \ref{Tab:Future}, we have simulated the impact of future data on our
fitted parameters. For this purpose we have kept the same central values of
the data points and reduced the errors according to the expected increase in
luminosity. Specifically, we have used that the $K_S\pi^-$ ($K^-\eta$) Belle
analysis employed $351$ ($490$) fb$^{-1}$ for a complete data sample of $1000$
fb$^{-1}$ accumulated at Belle-I for general purpose studies (we have assumed
the same resolution and efficiencies as in the published analyses following a
suggestion from the Collaboration). Similarly, we have also compared our
current results, eq.~(\ref{Final Results}), to the prospects for Belle-II at
the end of its data taking, with $50$ ab$^{-1}$ neglecting again possible
improvements in the detector response and data analysis. In the different
columns of Table \ref{Tab:Future}, we recall our results,
eq.~(\ref{Final Results}), and compare them, in turn, to the cases where both
decay modes are reanalysed using the whole Belle-I data sample, the same when
only one of the analysis is updated and analogously for Belle-II.

\begin{table}
\begin{center}
\begin{tabular}{|c|c|c|c|c|c|c|c|}
\hline
\backslashbox{\tiny{Error}}{\tiny{Data}}& \tiny{Current} & \tiny{Belle-I} & \tiny{Belle-I $K\pi$} & \tiny{Belle-I $K\eta$} & \tiny{Belle-II} & \tiny{Belle-II $K\pi$} & \tiny{Belle-II $K\eta$}\cr
\hline
\tiny{$\bar{B}_{K\pi}(\%)$} & \tiny{$0.404\pm0.012$} & \tiny{$\pm0.005$} & \tiny{$\pm0.005$} & \tiny{$\pm0.012$} & \tiny{$^\dagger$($0.001$)} & \tiny{$^\dagger$($0.001$)} & \tiny{$\pm0.012$}\cr
\tiny{$M_{K^*}$} & \tiny{$892.03\pm0.19$} & \tiny{$\pm0.09$} & \tiny{$\pm0.09$} & \tiny{$\pm0.19$} & \tiny{$^\dagger$($0.02$)} & \tiny{$^\dagger$($0.02$)} & \tiny{$\pm0.19$}\cr
\tiny{$\Gamma_{K^*}$} & \tiny{$46.18\pm0.44$} & \tiny{$\pm0.20$} & \tiny{$\pm0.20$} & \tiny{$\pm0.44$} & \tiny{$^\dagger$($0.02$)} & \tiny{$^\dagger$($0.03$)} & \tiny{$\pm0.42$}\cr
\tiny{$M_{K^{*\prime}}$} & \tiny{$1304\pm17$} & \tiny{$^\dagger$($7$)} & \tiny{$^\dagger$($9$)} & \tiny{$^\dagger$($8$)} & \tiny{$^\dagger$($1$)} & \tiny{$^\dagger$($1$)} & \tiny{$^\dagger$($1$)}\cr
\tiny{$\Gamma_{K^{*\prime}}$} & \tiny{$168\pm62$} & \tiny{$^\dagger$($19$)} & \tiny{$^\dagger$($24$)} & \tiny{$^\dagger$($25$)} & \tiny{$^\dagger$($3$)} & \tiny{$^\dagger$($4$)} & \tiny{$^\dagger$($11$)}\cr
\tiny{$\lambda^{\prime}_{K\pi}\times10^3$} & \tiny{$23.9\pm0.9$} & \tiny{$^\dagger$($0.3$)} & \tiny{$^\dagger$($0.3$)} & \tiny{$\pm0.8$} & \tiny{$^\dagger$($0.04$)} & \tiny{$^\dagger$($0.04$)} & \tiny{$\pm0.8$}\cr
\tiny{$\lambda^{\prime\prime}_{K\pi}\times10^4$} & \tiny{$11.8\pm0.2$} & \tiny{$\pm0.07$} & \tiny{$\pm0.07$} & \tiny{$\pm0.2$} & \tiny{$^\dagger$($0.01$)} & \tiny{$^\dagger$($0.01$)} & \tiny{$\pm0.2$}\cr
\tiny{$\bar{B}_{K\eta}\times10^4$} & \tiny{$1.58\pm0.10$} & \tiny{$\pm0.05$} & \tiny{$\pm0.10$} & \tiny{$\pm0.05$} & \tiny{$^\dagger$($0.01$)} & \tiny{$\pm0.10$} & \tiny{$^\dagger$($0.01$)}\cr
\tiny{$\gamma_{K\eta}(=\gamma_{K\pi})\times10^2$} & \tiny{$-3.3\pm1.3$} & \tiny{$^\dagger$($0.3$)} & \tiny{$^\dagger$($0.3$)} & \tiny{$^\dagger$($0.4$)} & \tiny{$^\dagger$($0.04$)} & \tiny{$^\dagger$($0.04$)} & \tiny{$^\circ$($0.3$)}\cr
\tiny{$\lambda^{\prime}_{K\eta}\times10^3$} & \tiny{$20.9\pm2.7$} & \tiny{$^\dagger$($0.7$)} & \tiny{$\pm2.7$} & \tiny{$^\dagger$($0.8$)} & \tiny{$^\dagger$($0.10$)} & \tiny{$\pm2.7$} & \tiny{$^\circ$($0.4$)}\cr
\tiny{$\lambda^{\prime\prime}_{K\eta}\times10^4$} & \tiny{$11.1\pm0.5$} & \tiny{$^\dagger$($0.2$)} & \tiny{$\pm0.5$} & \tiny{$^\dagger$($0.2$)} & \tiny{$^\dagger$($0.02$)} & \tiny{$\pm0.5$} & \tiny{$^\dagger$($0.06$)}\cr
\hline
\end{tabular}
\caption{\label{Tab:Future} \small{The errors of our final results
(\ref{Final Results}) are compared, in turn, to those achievable by analysing
the complete Belle-I data sample, and updating only the $K_S \pi ^-$ or
$K^-\eta$ analyses. The last three columns show the potential of fitting all
data collected by Belle-II and the same only for $K_S \pi ^-$ or for $K^-\eta$
(assuming the other mode has not been updated to include the complete Belle-I
data sample). Current Belle $K_S \pi ^-$ ($K^-\eta$) data correspond to $351$
($490$) fb$^{-1}$ for a complete data set of $\sim 1000$ fb$^{-1}=1$ ab$^{-1}$.
Expectations for Belle-II correspond to 50 ab$^{-1}$. All errors include both
statistical and systematic uncertainties. $^\dagger$ means that statistical
errors (in brackets) will become negligible, while $^\circ$ signals a tension
with the current reference best fit values. We thank Denis Epifanov for
conversations on these figures and on expected performance of Belle-II at the
detector and analysis levels. All errors have been symmetrised for
simplicity.}}
\end{center}
\end{table}

The majority of the expected errors for Belle-II will make completely 
negligible the statistical error with respect to the theoretical uncertainties, which then will most likely demand more elaborated approaches than those 
considered here. This would also happen in the case of the $K^*(1410)$ 
parameters with any updated Belle-I study. The impact of 
$\tau^-\to K^-\pi^0\nu_\tau$ on the $K^*(892)$ and $K^*(1410)$ meson parameters can be estimated by means of the $\tau^-\to K_S\pi^-\nu_\tau$ simulation. Such 
a measurement will be more significant in the determination of the $K^-\eta$ 
slope parameters than an updated study of this latter decay mode. In passing, 
we also mention that Belle-II statistics could be able to pinpoint 
possible inconsistencies between $\tau^-\to (K\pi)^-\nu_\tau$ and 
$\tau^-\to K^-\eta\nu_\tau$ data.

\section{Conclusions}\label{Concl}

Hadronic decays of the $\tau$ lepton remain to be an advantageous tool for
the investigation of the hadronization of QCD currents in the non-perturbative
regime of the strong interaction. In this work we have explored the benefits
of a combined analysis of the $\tau^-\to K_S\pi^-\nu_\tau$ and
$\tau^-\to K^-\eta\nu_\tau$ decays. This study was motivated by (our) separate
earlier works on the two decay modes considering them as independent data sets.
In particular, it was noticed in \cite{Escribano:2013bca} that the $K\eta$
decay channel was rather sensitive to the properties of the $K^*(1410)$
resonance as the higher-energy region is less suppressed by phase space.

Our description of the dominant vector form factor follows the work of
ref.~\cite{Boito:2008fq}, and proceeds in two stages. First, we write a
Breit-Wigner type representation (\ref{FpKpi2}) which also fulfils constraints
from $\chi$PT at low-energies. In eq.~(\ref{FpKpi2}), we have resummed the real
part of the loop function in the resonance denominators, but as was discussed
above, employing the following dispersive treatment, this is not really
essential. It mainly entails a shift in the unphysical mass and width
parameters $m_n$ and $\gamma_n$. Second, we extract the phase of the vector
form factor according to eq.~(\ref{del1Kpi2}) and plug it into the three-times
subtracted dispersive representation of eq.~(\ref{dispersive VFF}). This way,
the higher-energy region of the form factor, which is less well know, is
suppressed, and the form factor slopes emerge as subtraction constants of the
dispersion relation. A drawback of this description is that the form factor
does not automatically satisfy the expected $1/s$ fall-off at very large
energies. Still, in the region of the $\tau$ mass (and beyond), our form-factor
representation is a decreasing function such that the deficit should be
admissible without explicitly enforcing the short-distance constraint, thereby
leaving more freedom for the slope parameters to assume their physical values.

In our combined dispersive analysis of the $(K\pi)^-$ and $K^-\eta$ decays we
are currently limited by three facts: there are only published measurements of
the $K_S\pi^-$ spectrum (and not of the corresponding $K^-\pi^0$ channel), the
available $K^-\eta$ spectrum is not very precise and the corresponding data
are still convoluted with detector effects. The first restriction prevents us
from cleanly accessing isospin violations in the slope parameters of the vector
form factor. From our joint fits, we have however managed to get an indication
of this effect. The second one constitutes the present limitation in
determining the $K^*(1410)$ resonance parameters but one should be aware that
our approach to avoid the last one (assuming that the $K_S\pi^-$ unfolding
function gives a good approximation to the one for the $K^-\eta$ case) adds a
small (uncontrolled) uncertainty to our results that can only be fixed by a
dedicated study of detector resolution and efficiency. In this respect it would
be most beneficial, if unfolded measured spectra would be made available by
the experimental groups, together with the corresponding bin-to-bin correlation
matrices.

In Table~\ref{Tab:DifferentFits}, we have compared slightly different
options to implement constraints from isospin into the fits, and in
Table~\ref{Tab:ReferenceFit}, we studied the dependence of our fits on the
cut-off $s_{\rm cut}$ in the dispersion integral. Our reference fit is given
by the second column of Table~\ref{Tab:DifferentFits} and adding together the
statistical fit uncertainties with systematic errors from the variation of
$s_{\rm cut}$, our final results are summarised in eq.~(\ref{Final Results}).
The pole position we find for the $K^*(892)$ resonance is in perfect agreement
with previous studies. The main motivation of this work was, however, to
exploit the synergy of the $K\pi$ and $K\eta$ decay modes in characterising
the $K^*(1410)$ meson. According to our results, the relative weight $\gamma$
of both vector resonances is compatible in the $K\pi$ and $K\eta$ vector form
factors, which supports our assumption of their universality. With current
data we succeed in improving the determination of the $K^*(1410)$ pole mass,
but regarding the width, substantial uncertainties remain. Our central result
for these two quantities is
\begin{equation}\label{K^*'}
 M_{K^{*\prime}} \,=\, \left(1304 \pm 17\right)\,\mathrm{MeV} \,, \quad
\Gamma_{K^{*\prime}} \,=\,\left(171 \pm 62\right)\,\mathrm{MeV}\,,
\end{equation}
where we have symmetrised the uncertainties listed in eq.~(\ref{Final Results}).

We have then estimated the impact of future re-analyses including the complete
Belle-I data sample and all expected data from Belle-II on these decay modes.
This projection reveals (in both cases) that the increased statistics will
most probably require a refined theoretical framework to match the experimental
precision in the determination of the $K^*(1410)$ resonance parameters.
While our description so far is purely elastic, this may include incorporation
of coupled channels to take into account inelastic effects along the lines of
refs.~\cite{Moussallam:2007qc,Bernard:2013jxa}, which would allow for a proper
inclusion of higher channels in the resonance widths. Belle-II data would also
lead to much improved tests of our low-energy description and the $K^*(892)$
dominance region. Knowledge of isospin breaking effects on the slope parameters
could be drastically improved by measuring the hadronic invariant mass
distribution in $\tau^-\to K^-\pi^0\nu_\tau$ decays, which would by the way
increase the accuracy in the extraction of the $K^*(892)$ pole position. We
hope that this study will give additional motivation to the B-factory
collaborations for performing the respective analyses.

\appendix
\section{Exponential parametrisation of the vector form factor}
\label{AppA}

The exponential parametrisation of $f_+^{K\pi}(s)$ is a variant of the form
factor Ansatz (\ref{FpKpi2}) in which the real part of $\tilde H_{K\pi}(s)$
is resummed into an exponential function
\cite{Jamin:2006tk,Jamin:2008qg,Guerrero:1997ku},
\begin{equation}
\label{VFFexp}
f_+^{K\pi}(s) \,=\, \left[\,
\frac{m_{K^*}^2+\gamma s}{D(m_{K^*},\gamma_{K^*})}
-\frac{\gamma s}{D(m_{K^{*\prime}},\gamma_{K^{*\prime}})}
\,\right]{\rm e}^{\frac{3}{2}{\rm Re}\widetilde H_{K\pi}(s)} \,,
\end{equation}
where now $D(m_n,\gamma_n) \,=\, m_n^2 - s - i\,m_n\gamma_n(s)$ and the
energy-dependent resonance widths, defined as
\begin{equation}
\label{widthexp}
\gamma_n(s) \,=\, \gamma_n\,\frac{s}{m_n^2}\,
\frac{\sigma_{K\pi}^3(s)}{\sigma_{K\pi}^3(m_n^2)}\,,
\end{equation}
are equal to the imaginary part of the propagator in eq.~(\ref{Dden}) through
the identification $\kappa_n\,{\rm Im}\widetilde H_{K\pi}(s)=m_n\gamma_n(s)$.
This representation of $f_+^{K\pi}(s)$ in the elastic limit was used beyond
this approximation in refs.~\cite{Jamin:2006tk,Jamin:2008qg} including the
$K\eta$ channel and ref.~\cite{Escribano:2013bca} also incorporating the
$K\eta'$ effects. However, in order to perform a fair comparison of the results
obtained from this parametrisation and the dispersive representation in
eq.~(\ref{dispersive VFF}) we work in the elastic limit and use for
$\widetilde H_{K\pi}(s)$ the isospin average of eq.~(\ref{HtKpi}). Needless to
say, the unphysical ``mass'' and ``width''  parameters $m_n$ and $\gamma_n$
in this parametrisation will be different from their analogues in the
dispersive treatment but the corresponding pole parameters should not differ
significantly. It is worth mentioning, however, that when the normalised
version of the form factor in eq.~(\ref{VFFexp}) is directly confronted with
experimental data the slope parameters are not fitted but deduced from the
Taylor expansion of the form factor (unlike the test proposed in the main text
where the phase of the form factor is calculated first and then inserted into
the dispersive relation).

In Table \ref{Tab:ExpFit}, we display the results of the direct application of
the exponential vector form factor in eq.~(\ref{VFFexp}) using three different
settings: a combined fit of the two sets of data with
$\gamma_{K\pi}=\gamma_{K\eta}$ (Fit I, which implies
$\lambda_{K\pi}^{\prime(\prime)}=\lambda_{K\eta}^{\prime(\prime)}$); the same
but $\gamma_{K\pi}\neq\gamma_{K\eta}$ (Fit II); and fitting the data sets
separately (Fit III). In the last case, the pole position of the $K^*(892)$
resonance is obtained from the fit to $K\pi$ data and then plugged into the
$K\eta$ fit. On the contrary, the $K^*(1410)$ pole position is kept free in
both fits (in brackets the results from the fit to $K\eta$ data alone). Looking
at the various $\chi^2/$n.d.f.~of Table \ref{Tab:ExpFit}, one immediately
realises the meagre performance exhibited by the exponential parametrisation
as compared to the dispersive representation achievements shown in
Table~\ref{Tab:DifferentFits}. In the $K\eta$ part of Fit III (fourth column)
the $\chi^2/$n.d.f.$\,\sim 2$. Particularly inept are the values obtained for
the $K\eta$ branching ratio which are in all cases far from the experimental
measurement. Therefore, a combined analysis of the $\tau^-\to K_S\pi^-\nu_\tau$
and $K^-\eta\nu_\tau$ decays clearly disfavours the direct exponential
treatment as compared to the dispersive approach, a conclusion which was
already hinted at by the independent analysis of $K\eta$ data in
ref.~\cite{Escribano:2013bca}. Now comparing, for instance, Fit II in
Table~\ref{Tab:ExpFit} with its analogue Fit~B in Table~\ref{Tab:DifferentFits},
it is seen that the pole positions of both resonances are quite in agreement in
the two approximations as also happens with their relative weights. However,
somewhat larger values with smaller errors are obtained for all the different
slope parameters, in accord this time with the previous analyses
in refs.~\cite{Jamin:2006tk,Jamin:2008qg}.

\begin{table}
\begin{center}
\begin{tabular}{|c|c|c||c|}
\hline
Fitted value& Fit I & Fit II & Fit III \cr
\hline
$\bar{B}_{K\pi}(\%)$ & $0.394\pm0.008$ & $0.398\pm0.009$ & $0.401\pm0.009$\cr
$(B_{K\pi}^{th})(\%)$ & $(0.391)$ & $(0.394)$ & $(0.398)$\cr
$M_{K^*}$ & $892.35\pm0.25$ & $892.31\pm0.25$ & $892.39\pm0.23$\cr
$\Gamma_{K^*}$ & $47.19\pm0.51$ & $47.21\pm0.49$ & $47.15\pm0.46$\cr
$M_{K^{*\prime}}$ & $1318\pm10$ & $1318\pm11$ & $1265\pm16\,(1340\pm19)$ \cr
$\Gamma_{K^{*\prime}}$ & $146\pm31$ &$165\pm36$ &$145\pm42\,(218\pm65)$\cr
$\gamma_{K\pi}\times10^2$ & $=\gamma_{K\eta}$ & $-4.1\pm0.9$ & $-3.8\pm1.0$\cr
$\lambda^{\prime}_{K\pi}\times10^3$ & $25.02\pm0.13$ & $25.08\pm0.14$ & $25.16\pm0.14$\cr
$\lambda^{\prime\prime}_{K\pi}\times10^4$ & $12.56\pm0.10$ & $12.61\pm0.10$ & $12.66\pm0.11$\cr
$\bar{B}_{K\eta}\times10^4$ & $1.34\pm0.07$ & $1.35\pm0.08$ & $1.25\pm0.11$\cr
$(B_{K\eta}^{th})\times10^4$ & $(1.15)$ & $(1.16)$ & $(1.06)$\cr
$\gamma_{K\eta}\times10^2$ & $-4.6\pm0.8$ & $-6.2\pm1.6$ & $-8.4\pm2.7$\cr
$\lambda^{\prime}_{K\eta}\times10^3$ & $=\lambda^{\prime}_{K\pi}$ & $24.80\pm0.23$ & $24.47\pm0.40$\cr
$\lambda^{\prime\prime}_{K\eta}\times10^4$ & $=\lambda^{\prime\prime}_{K\pi}$ & $12.40\pm0.17$ & $12.18\pm0.29$\cr
$\chi^2/$n.d.f. & $188.4/109\sim1.72$ & $184.0/108\sim1.70$ & $(117.9+49.5)/(81+25)\sim1.58$\cr
\hline
\end{tabular}
\caption{\label{Tab:ExpFit}
\small{Fit results obtained using the exponential parametrisation for
different settings: a combined fit of $K\pi$ and $K\eta$ data with
$\gamma_{K\pi}=\gamma_{K\eta}$ (Fit~I), the same but
$\gamma_{K\pi}\neq\gamma_{K\eta}$ (Fit~II); and fitting the data separately
(Fit~III). See the main text for further details. Dimensionful parameters are
given in MeV. As a consistency check, for each of the fits we provide (in
brackets) the value of the respective branching ratios obtained by integrating
eq.~(\ref{spectral function})}}
\end{center}
\end{table}

\acknowledgments
We are indebted to Denis~Epifanov and Simon~Eidelman for discussions on
the Belle analysis and the prospects for Belle-II. We appreciate very much
correspondence with Swagato~Banerjee and Ian~Nugent regarding the BaBar studies.
This work was supported in part by the FPI scholarship BES-2012-055371 (S.G-S),
the Ministerio de Ciencia e Innovaci\'on under grant FPA2011-25948,
the Secretaria d'Universitats i Recerca del Departament d'Economia i
Coneixement de la Generalitat de Catalunya under grant 2014 SGR 1450,
the Ministerio de Econom\'{\i}a y Competitividad under grant SEV-2012-0234,
the Spanish Consolider-Ingenio 2010 Programme CPAN (CSD2007-00042), and
the European Commission under programme FP7-INFRASTRUCTURES-2011-1
(Grant Agreement N. 283286). P.R.~acknowledges funding from CONACYT and DGAPA
through project PAPIIT IN106913.

% The bibliography will probably be heavily edited during typesetting.
% We'll parse it and, using the arxiv number or the journal data, will
% query inspire, trying to verify the data (this will probably spot
% eventual typos) and retrieve the document DOI and eventual errata.
% We however suggest to always provide author, title and journal data:
% in short all the informations that clearly identify a document.

\end{document}